# Application of the renormalization group method in wireless market intelligence

M.V. Simkin[1] and J. Olness[2]

*Telephia, Inc., 1 Beach St., San Francisco, CA 94133*

We use a renormalization group method, similar to that developed for random spin chains, to infer information about the layouts of cellular wireless networks.

Methods developed for statistical mechanics are occasionally found to be useful in practical, 'real-world' applications. Perhaps the best-known example is the Simulated Annealing method [1], which was originally developed to find the ground states of the Sherrington-Kirkpatrick model [2] of spin glasses. This method is now used for a number of commercial applications (including optimization of cellular wireless networks [3]).

This letter applies a real-space renormalization group technique to the real-world problem of inferring the cell locations of cellular wireless networks, a valued piece of market intelligence. (The method is similar to that of Ma, Dasgupta and Hu [4], and Fisher [5], which were developed to study spin glasses.)

In order to serve a large number of wireless phone users with a limited amount of radio spectrum, wireless network operators divide their service areas into a large number of sub-regions, which are referred to as 'cells'. Each cell is served by its own antenna, and has use of a portion of the total spectrum available to the network. In order to keep interference to manageable levels, nearby cells do not use the same piece of spectrum. Provided they are far enough away, however, distant cells can 'reuse' the same portion of spectrum, thereby enabling the wireless network to serve a very large number of simultaneous users. (For an excellent introduction to cellular wireless networks, see Ref. [6].)

In order to optimize their networks and to benchmark themselves against competitors, wireless network operators frequently test their networks using mobile test equipment. During the course of a so-called 'drive-test', wireless telephone calls are placed from many locations throughout the network. The test equipment records a large number of statistics associated with each call, including time, location, signal strength, bit-error rate, and signal-to-noise ratio. The results are used both to understand the end-user experience and to diagnose performance problems.

---

[1] Email: msimkin@telephia.com
[2] Email: jolness@yahoo.com

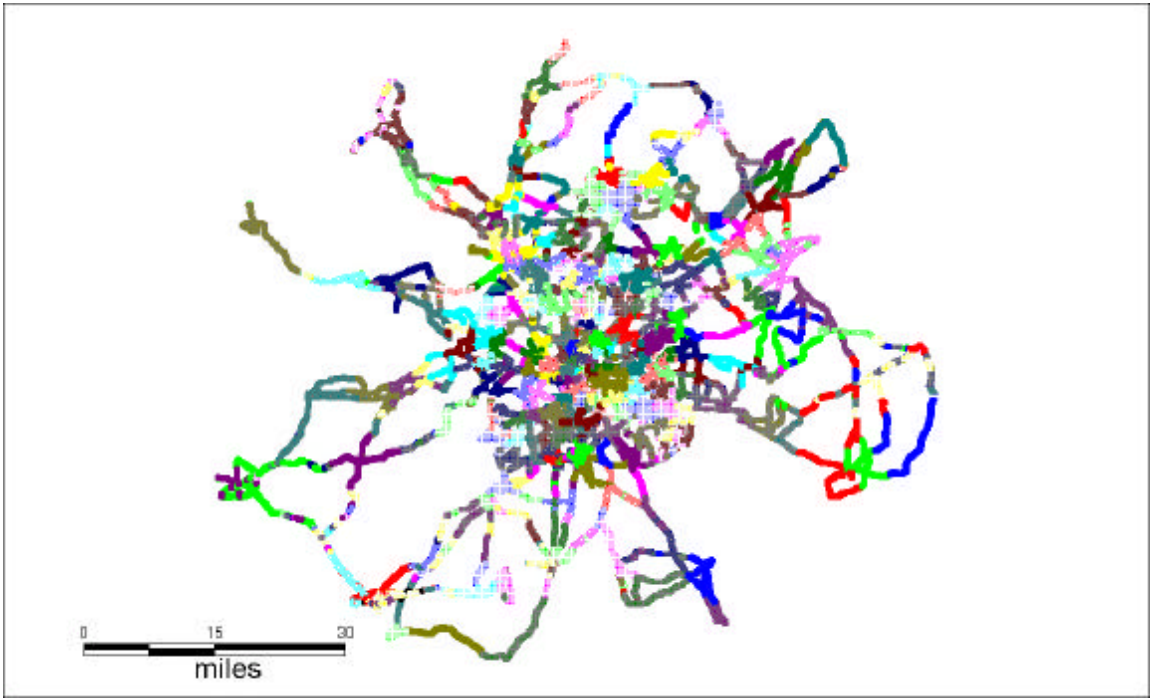

**Figure 1** GPS points colored according to 'color-code' for a major wireless carrier in Atlanta. Each distinct collection of colored points indicates the approximate location and extent of a cell.

One of the goals of a drive-test is to gain insight about competitor networks. Until now, this insight has been limited to simple, 'quality-of-service' comparisons: "At this location, how does my network stack up in terms of network availability, audio quality, and call-success rate?" It would be valuable to know how competitors have laid out their networks, and in particular, where they have placed their transmitters. We have discovered a method of estimating the number and locations of antennas in any common wireless network

The method relies upon what will be referred to as a color-code. Every transmission from the wireless network contains a color-code, which identifies the antenna from which the transmission originated [7]. (Color-codes are called by different names depending on the type of network, but all common wireless voice networks employ such codes [8].) Every measurement recorded by the mobile test equipment includes the GPS (latitude and longitude) coordinates of the location at which the measurement was made. In principle, antenna locations can be approximated simply by averaging the GPS coordinates that are associated with each color-code. (Figure 1 shows GPS points colored according to color-code for a major wireless network in Atlanta [9]. The approximate cell locations are already apparent simply by inspecting the diagram.)

Unfortunately, it is common practice to reuse color-codes within a network. Intended primarily as a means of reducing mutual interference effects, color-codes can safely be reused for antennas that are sufficiently far away from one another. In other words, color-codes identify, not individual antennas, but whole families of antennas. This presents an obvious problem for the naïve methodology outlined in the preceding paragraph: How can we distinguish measurements associated with one (unknown) antenna location from those associated with other antennas belonging to the same family?

Since antennas sharing a common color-code will generally be widely separated from one another, this challenge presumably could be overcome using simple visual inspection. In practice, this proves to be a complex problem. Wireless networks typically contain hundreds of antennas, which effectively rules out a 'manual' solution; an automated solution is called for. The goal of this paper is to describe a solution based on the renormalization group method.

In order to motivate the technique that we developed, it is instructive to describe a naïve (and ineffective) alternative. First specify a cutoff distance $r$; this is the maximum allowable cell radius in the model. Then select a random 'seed' point from the drive-test data. All data points having the same color-code and lying within distance $r$ of the seed point are assigned to a single cell. A new seed point is then selected from the remaining data points and the process is repeated until there are no more data points.

The problem with this approach is that the random seed points are not guaranteed to lie near the centers of cells. Due to the vagaries of radio wave propagation, signals from a given cell can occasionally be picked up at locations that are in fact closer to another cell in the same family (i.e., having the same color-code). If one of these points is selected as a seed, the result is likely to be that either the two nearby cells are combined inappropriately into a single cell, or that a new 'phantom' cell is created by robbing data points from the two nearby cells. Whatever the case, it is clear that outlying points pose a significant challenge. Fortunately, renormalization group techniques can be used to ensure that these points are not selected as seeds.

The idea of the renormalization group is to systematically get rid of short length degrees of freedom in order to find essential features on large length scale. This is similar to what we need: find clusters of points, which are seen when we do not distinguish any distances smaller then color-code reuse distance. However, our system is not on a perfect lattice: distances between GPS points are random. Therefore an implementation of the RG for disordered system would be appropriate. Such implementation was developed in Refs 4-5 for random spin chains. The main idea is to first cluster the pair of spins with the strongest coupling, obtain the new effective Hamiltonian, and repeat the procedure *ad physicum*. In random spin chains randomness is in energy. In our problem randomness is in space. Therefore we would cluster a pair of points with minimum separation, and repeat the procedure with resulting set of points until the minimum separation between points exceeds the color-code reuse distance.

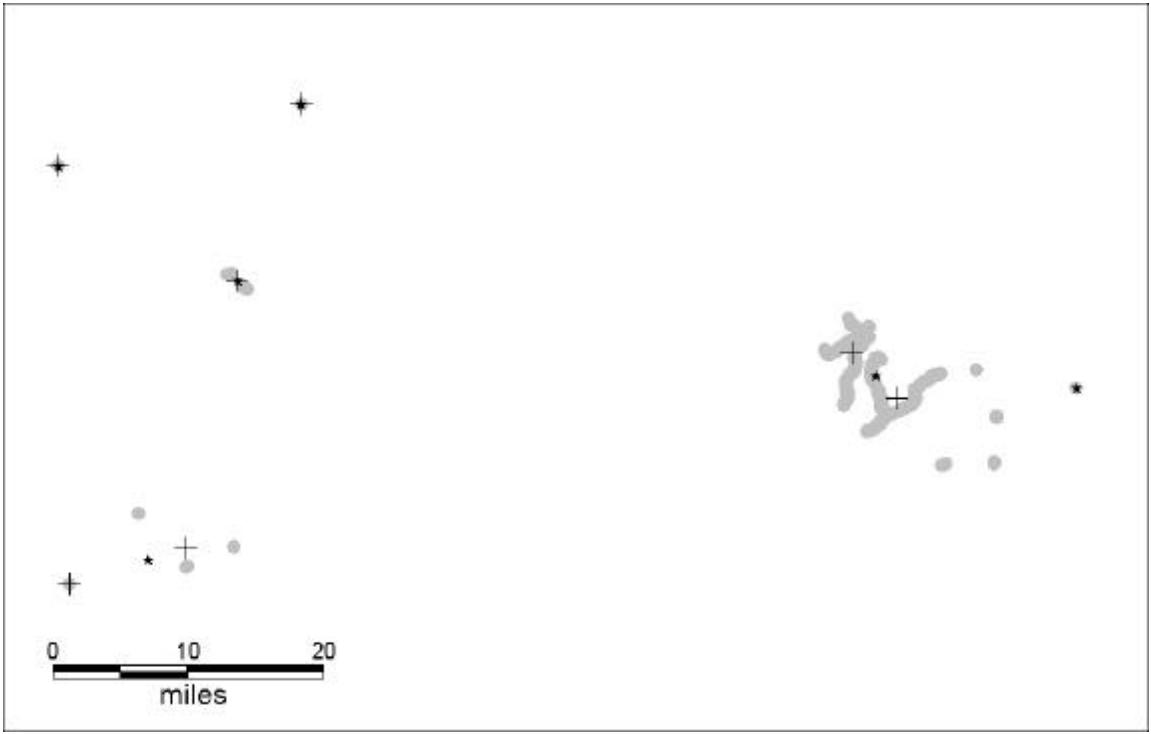

**Figure 2** Gray circles are GPS points with one particular color-code value. Crosses are the cell centers computed using original naive approach. Stars are cell centers computed using the renormalization group.

The RG prescription for clustering data points is as follows: First choose a color-code and select all data points having that color-code. Assign a weight factor, $n$, to each data point; initially, $n = 1$ for all data points. Compute the distance between each pair of data points. Select the pair of points, $i$ and $j$, having the smallest separation distance and combine them using the following formulas:

$$n = n_i + n_j; x = \frac{n_i x_i + n_j x_j}{n_i + n_j}; y = \frac{n_i y_i + n_j y_j}{n_i + n_j}.$$

Repeat this procedure until the minimum separation distance exceeds the cutoff distance.

Figure 2 shows the results for a representative color-code. The gray dots represent the original data points and the stars represent the output of the algorithm. For comparison purposes, the results of the naïve method are included as well, and are shown as crosses.

The points on the right-hand side of the page indicate an obvious failure of the naïve method. The dense cluster of points is almost certainly associated with a single cell, yet the naïve method has identified two distinct clusters. The centroids of the two clusters are separated by a distance of only 5 miles, which is substantially less than the cutoff distance of 12 miles; this is certainly not what we had in mind when we introduced the cutoff. (The RG method also identified a second cluster, but the second 'cluster' consists of a single point whose distance from the main cluster is greater than the cutoff distance. The way to cure this is to eliminate 'clusters' consisting of only few points.)

The naïve method and the RG method can be compared systematically by studying the rank-order of cell-center separation distances. If a given cell-center is found to be among, say, the ten nearest neighbors of another cell-center having the same color-code, we should suspect that the method has erroneously 'split' a single cluster of data points, identifying a pair of clusters where there should be only one.

Figure 3 presents the results of such an analysis for both the naïve method and the RG method. The horizontal axis indicates the rank-order of the nearest cell having the same color-code, and the vertical axis indicates the frequency with which each value occurs. The naïve method has a tendency to split cells inappropriately. This shows up in the histogram as a significant instance of cases where the rank associated with a given cell-center is less than, say, ten. By comparison, the smallest rank associated with any cell-center identified by the RG method was 64.

It is easy to show that if there are $N$ color-codes, then the nearest cells having the same color-code should be about $pN$'s nearest neighbor. Since we did not sample all of the cells in the network, we expect actual number from our data to be somewhat lower than the theoretical figure. This agrees with the maximum at around 512 for RG data seen on the histogram of Figure 3 (the $N$ in our case is 256), and disagrees with the maximum at 256 for the original method.

The cell centers found using the renormalization group approach are shown in Figure 4. (The radius of each circle is defined to be half the distance to its nearest neighbor. This serves to indicate the approximate cell size and location of the cell, and ensures that none of the circles overlaps any of its neighbors.)

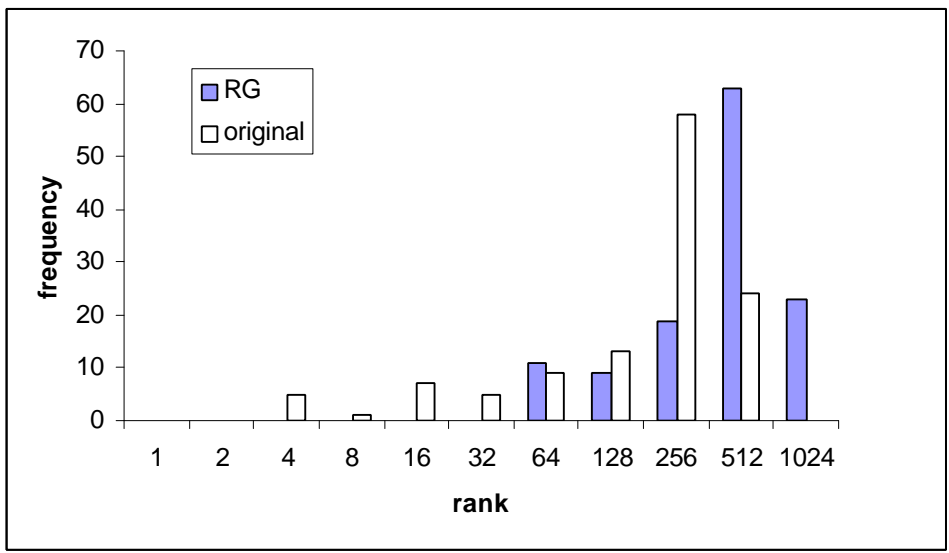

**Figure 3** Rank-histogram of the nearest cell having the same color-code for original naïve method and RG method.

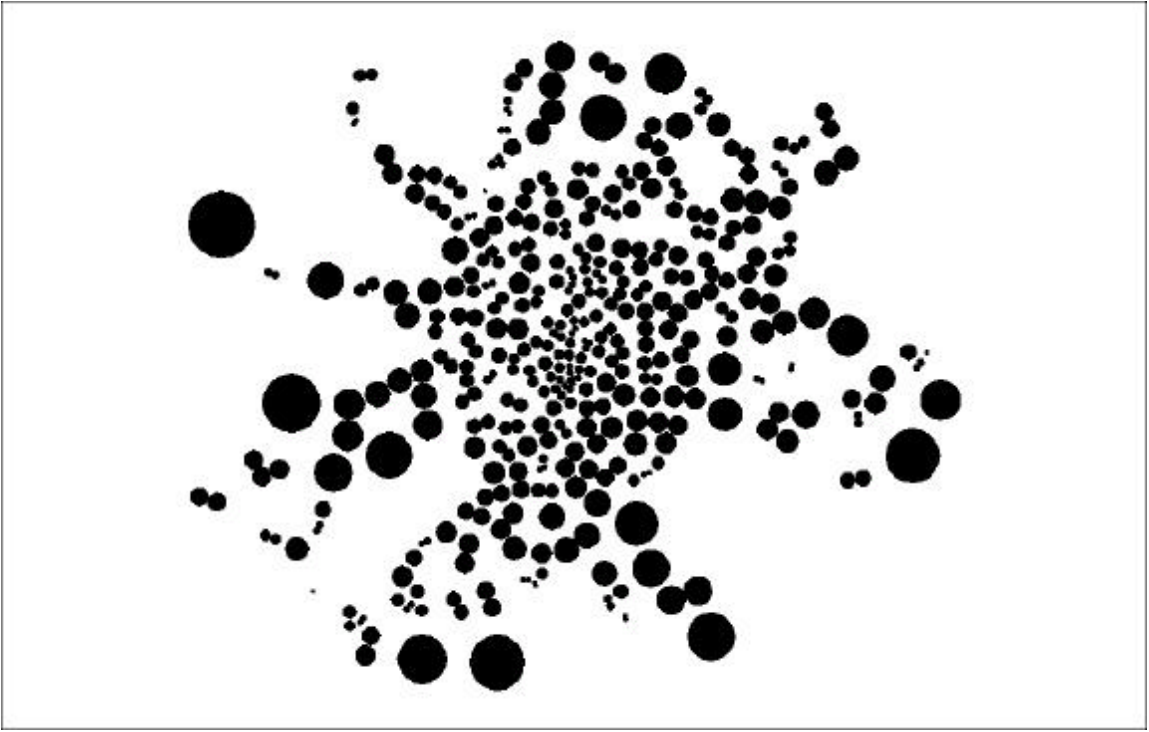

**Figure 4** Approximate cell locations for a major wireless network in Atlanta. For display purposes, the radius of each circle is defined to be half the distance to the nearest-neighbor cell.

In conclusion, we have successfully applied renormalization group techniques to the commercial problem of gathering intelligence about the layouts of cellular wireless networks.